\documentclass[fleqn,usenatbib]{mnras}
\usepackage{comment}
\usepackage{newtxtext,newtxmath}

\usepackage[T1]{fontenc}
\usepackage{adjustbox}
\usepackage{float}
\usepackage{caption}

\DeclareRobustCommand{\VAN}[3]{#2}
\let\VANthebibliography\thebibliography
\def\thebibliography{\DeclareRobustCommand{\VAN}[3]{##3}\VANthebibliography}

\usepackage{graphicx}	
\usepackage{amsmath}	
\usepackage{xcolor}

\title[ComPACT: ACT+Planck galaxy cluster catalogue]{ComPACT: combined ACT+Planck galaxy cluster catalogue}

\author[S.Voskresenskaia et al.]{
S.Voskresenskaia,$^{1,2}$\thanks{E-mail: savoskresenskaya@edu.hse.ru}
A.Meshcheryakov,$^{2,3}$\thanks{E-mail: mesch@cosmos.ru}
N.Lyskova$^{2,4}$
\\
$^{1}$Higher School of Economics, Moscow, Russia\\
$^{2}$Space Research Institute RAS, Moscow, Russia\\
$^3$Faculty of Computational Mathematics and Cybernetics, Lomonosov Moscow State University, Russia\\
$^4$Astro Space Center, Lebedev Physical Institute, Russian Academy of Sciences, Russia\\
}

\pubyear{2023}

\begin{document}
\label{firstpage}
\pagerange{\pageref{firstpage}--\pageref{lastpage}}
\maketitle

\begin{abstract}
Galaxy clusters are the most massive gravitationally bound systems consisting of dark matter, hot baryonic gas and stars. They play an important role in observational cosmology and galaxy evolution studies. We develop a deep learning model for segmentation of Sunyaev-Zeldovich (SZ) signal on ACT+Planck intensity maps and construct a pipeline for microwave cluster detection in the ACT footprint. The proposed model allows us to identify previously unknown galaxy clusters, i.e. it is capable of detecting SZ sources below  the detection threshold adopted in the published galaxy clusters catalogues (such as ACT DR5 and PSZ2). In this paper, we use the derived SZ signal map to considerably improve a cluster purity in the extended catalogue of Sunyaev-Zeldovich objects from Planck data (SZcat) in the ACT footprint. From SZcat, we create a new microwave galaxy cluster catalogue (ComPACT), which includes 2,962 SZ objects with cluster purity conservatively estimated as $\gtrsim74-84$\%. We categorise objects in the catalogue into 3 categories, based on their cluster reliability. Within the ComPACT catalogue, there are $\gtrsim{}977$ new clusters with respect to the ACT DR5 and PSZ2 catalogues.
\end{abstract}


\section{Introduction}
\label{sec:intro}
\thispagestyle{empty}
Galaxy clusters are the most massive gravitationally bound systems in the Universe \citep[see][for reviews]{Clusters,2012ARA&A..50..353K}. These objects are located at the nodes of the cosmic web and serve as reliable tracers of the underlying dark matter distribution. The cluster mass function, which describes the number density of galaxy clusters as a function of their mass, and its evolution with redshift provide a critical test of structure formation models and allows to constrain cosmological parameters \citep[e.g.][among others]{2009ApJ...692.1060V, Mass_scale:Pratt_2019, 2012AstL...38..347B,2015MNRAS.446.2205M}.

Galaxy clusters are composed of dark matter, which makes up around 85\% of their total mass, the intracluster medium (ICM; $\sim$ 12\% of the total mass), which is an ionised hot gas, and stars ($\sim$ 3\%). Consequently, they can be detected at several wavelength bands. In the optical range, galaxy clusters represent a concentration of elliptical galaxies. Clusters are also bright in X-rays mainly due to thermal bremsstrahlung and line emission produced by the ionised ICM \citep{1988xrec.book.....S}. 
X-ray catalogues may provide a wealth of information about galaxy clusters, such as their temperature, luminosity, and gas mass, which are essential for understanding the formation and evolution of these structures. The same hot ICM (namely, the high-energy electrons) also interacts with low-energy cosmic microwave background (CMB) radiation through the inverse Compton scattering creating a distortion in the black-body spectrum (the thermal Sunyaev-Zeldovich effect; \cite{1970Ap&SS...7....3S, 1972CoASP...4..173S}). The ICM boosts CMB photons to higher frequencies resulting in a characteristic increase/decrease in the CMB intensity depending on the specific microwave frequency in which a galaxy cluster is being observed. Studying galaxy clusters in (sub)millimetre range via the SZ effect has several advantages over other wavelength ranges. First, the magnitude of the SZ effect  does not depend on redshift, thus providing a powerful tool for detecting clusters at high redshifts. Second, the SZ signal is proportional to the integrated gas pressure along the line of sight through the cluster, which is directly linked to the total mass of a cluster. The relationship between the integrated Compton parameter $Y_{SZ}$ (the total SZ flux) and a cluster mass $M$ has already proved to be particularly tight \citep[see ][among many others]{2006ApJ...650..128K, 2013A&A...550A.129P}. Although, a scatter in the $Y_{SZ}$-mass relation  is not negligible since clusters are not spherical, they have different accretion histories, a contribution of the non-thermal pressure support in cluster outskirts may be significant \cite[e.g.][]{2012ApJ...758...74B, 2015ApJ...807...12Y, 2020MNRAS.496.2743G}. Moreover, the presence of non-thermal pressure in the cluster introduces a bias in mass determination from the SZ (or X-ray) observations as  masses are typically estimated under the assumption of the hydrostatic equilibrium between the gravitational potential and the observed thermal (not total) pressure \cite[][among many others]{2007ApJ...655...98N, 2013ApJ...777..123B,2016ApJ...827..112B, 2012NJPh...14e5018R}.The third advantage of studying clusters via the SZ effect is the ability to build a virtually mass-limited sample. SZ selected clusters are shown to be less biased in favor of cool-core and/or relaxed systems compared to X-ray selected samples \citep[e.g.][]{2011A&A...526A..79E, 2016MNRAS.457.4515R, 2017ApJ...846...51L}. 

The number of SZ catalogues have been released in recent years including the Planck survey (\cite{Planck:14, Planck:16b}), the South Pole Telescope (SPT, \cite{Bleem_2015} and the Atacama Cosmology Telescope (ACT; \cite{Hilton_2021}). With ongoing and upcoming CMB surveys, more than $10^4$ galaxy clusters are expected to be detected \citep{2022ApJ...926..172R}. Together with reliable mass estimates\footnote{In practice, however, we cannot  observe a cluster mass directly. Thus, we have to rely on mass proxy estimates that can be (and they are) scattered and biased. To date, cluster masses derived from the weak lensing analysis are considered to be less biased compared to other approaches. \cite{2023arXiv230200687E} showed that  weak lensing masses tend to be, on average, biased low by 5–10\% with respect to the true mass. While masses derived from X-ray observations under an assumption of a hydrostatic equilibrium are biased low with respect to the true mass by around 20-25 \% \citep[][among many others]{2021MNRAS.502.5115G}. A more detailed discussion is given, for example, in   \cite{2014A&A...571A..20P}, Appendix A.}, such cluster samples may have a ground-breaking impact on our understanding of the structure growth in the Universe,  the dark energy equation of state, and modified  gravity theories \citep[e.g.][]{2015ApJ...799..214B,  2016A&A...594A..13P, 2012AstL...38..347B, 2015MNRAS.446.2205M, 2018AstL...44..653B, 2013AstL...39..357B} .

Recently, \cite{SZcat} presented SZcat -- a catalogue of Planck extended SZ sources with a low level of significance (so that nearly all possible Planck cluster candidates are there along with a large number of spurious detections). Here, we aim to develop a method for an automated cluster detection on publicly available combined ACT and Planck maps \citep{Naess_2020} and to apply it to the fields of SZcat sources. As a result, we obtain a new catalogue of galaxy clusters -- ComPACT.

The paper is organised as follows. In the next section, we review deep learning models for SZ clusters detection proposed in literature. Section~\ref{sec:dataset} describes data used in this work. In Section~\ref{sec:method}, we describe a deep learning model for cluster segmentation and a procedure for an object detection. Then, in Section~\ref{sec:compact}, the ComPACT catalogue is described. In Section~\ref{sec:conclusions}, our conclusions are presented.   

\subsection{Deep learning models for SZ clusters detection}
In the last few years, deep learning (DL) approaches based
mostly on CNN (Convolution Neural Network) architectures were successfully applied to various object detection and segmentation problems in observational astronomy in different spectral domains: e.g. to imaging data in radio \citep{2023MNRAS.523.1967H}, microwave \citep{SZcat,2021MNRAS.507.4149L,Bonjean_2020}, and optical \citep{2019MNRAS.490.3952B} spectral ranges. The key advantages of deep learning models over classical astronomical object detection methods (such as SExtractor \citep{1996AAS..117..393B} in the optical or Matched Multi-filter approach (MMF, \cite{MMF_Melin:06, MMF_Williamson:11}) in the microwave) are: (i)  unification of an object detection model architecture over many spectral domains, and (ii) the amazing ability of a DL method to improve itself with the increase of size of available training sample. As it was shown in SKA Science Data Challenge 2 \citep{2023MNRAS.523.1967H}, deep learning models trained on a large and representative enough knowledge base outperform a classical approaches in the astronomical object detection tasks.

As mentioned above, DL approaches were successfully applied to microwave imaging data. \citet{Bonjean_2020} proposed to use a DL model with the encoder-decoder architecture to make a segmentation map of SZ signal from Planck HFI imaging data and detected SZ sources on it. \cite{2021AstBu..76..123V} trained a CNN classification model to recognise SZ sources in Planck intensity maps. \cite{2021MNRAS.507.4149L} proposed a hybrid model DeepSZ, based on a combination of CNN and a classical MMF approach and demonstrated its power in detecting SZ sources using the simulated CMB maps (with characteristics similar to SPT-3G survey). 

In all papers mentioned above no catalogue of SZ sources was published. Recently, \cite{SZcat} trained a U-Net model on Planck HFI data and presented SZcat -- a full catalogue of cluster candidates. SZcat is a large combined sample of galaxy cluster candidates (up to a low confidence threshold) selected by two methods: (a) using the DL model (described in \cite{SZcat}) and applying it to Planck HFI intensity maps; and by (b) using a classical detection approach \citep{2017AstL...43..507B} and applying it to Compton parameter maps published in Planck 2015 data release \citep{2016AA...594A..24P}. By its construction, SZcat is characterized by high completeness and low cluster purity of approximately 22\%. We use the SZcat catalogue further in our work.

Another approach to detect new objects is to combine data from different telescopes. It makes possible to lower the threshold for cluster detection and discover previously undetected galaxy clusters. For example, \citet{PSZSPT} combined Planck and SPT microwave data; \citet{ComPRASS} combined Planck data with ROSAT X-ray catalogue. In both cases of Planck-SPT and Planck-ROSAT data combination, new cluster catalogues were published (with higher completeness/purity than cluster catalogues obtained from Planck/SPT/ROSAT data individually).

\section{Data}
\label{sec:dataset}
\thispagestyle{empty}
We use publicly available composite ACT+Planck intensity maps\footnote{\url{https://lambda.gsfc.nasa.gov/product/act/actpol_dr5_coadd_maps_get.html}} from \cite{Naess_2020} at 90, 150, and 220 GHz with 0.5 arcmin per pix resolution covering around 18,000 squared degrees of the sky area (which we call as the ACT footprint hereafter). To combine ACT and Planck data, \cite{Naess_2020} used the HFI Planck maps at 100, 143 and 217 GHz, i.e. at frequencies close to the ACT ones. Authors split maps into tiles and model the tiles as noisy, transformed versions of a single underlying sky. They then used the overlap to seamlessly merge the tiles into a single map.

In order to train and test a SZ cluster segmentation model (see Section \ref{sec:method}), we use the following catalogues of galaxy clusters and radio sources found in the ACT data: 
\begin{itemize}
    \item the ACT DR5 cluster catalogue \citep{Hilton_2021}, which contains 4,195 optically confirmed galaxy clusters. The sources were selected by using the MMF method for the ACT multi-frequency data collected from 2008 to 2018. In Section~\ref{subsec:inters}, we illustrate the mass-redshift dependence for ACT clusters using the cluster masses from the M500cUncorr column \citep[see Table 1 in][]{Hilton_2021}, which were obtained by assuming the universal pressure profile and the scaling relation between the integrated Compton parameter $Y_{SZ}$ and $M_{500}$ (a halo mass contained within a radius  inside of which the mean interior density is 500 times the critical density) from \cite{2010A&A...517A..92A};
    \item a catalogue of point sources (predominantly, active galactic nuclei (AGN)) from \cite{2019MNRAS.486.5239D}, which covers 680 squared degrees of the full ACT footprint;
    \item the equatorial catalogue of extragalactic sources \citep{Gralla_2020}, which contains 287 dusty star-forming galaxies (DSFG) and 510 radio-loud active galactic nuclei.
 \end{itemize}   

Additionally, for training and testing our classification model (see Sections \ref{subsec:train_test_sample}--\ref{subsec:test_sample}), we select a number of random fields which do not contain neither ACT DR5 clusters nor radio sources from  \cite{2019MNRAS.486.5239D, Gralla_2020}. Since the density of clusters on the sky is low (see our estimates in Section~\ref{subsec:test_sample}), these fields are expected to be free of galaxy clusters. 

We construct a new catalogue of galaxy clusters on the basis of the extended Planck cluster candidates sample SZcat, which is expected to contain almost all microwave clusters seen in the Planck data. In the ACT footprint (which we consider here), the total number of SZcat sources is $N_{SZcat}=14,850$.

To estimate the purity and completeness of cluster detections in a resulting catalogue (see Section~\ref{sec:compact}), we have selected $N_{field}=14,850$ random directions on the sky in the ACT footprint. 

For test purposes, we cross-correlate a new catalogue with the following external X-ray and SZ cluster catalogues: MCXC, 4XMM DR12, SPT-SZ, ComPRASS, and PSZSPT (see Appendix\,\ref{sec:appB}).  

\section{Method}
\label{sec:method}
\begin{figure*}
    \centering
    \includegraphics[width=1\linewidth]{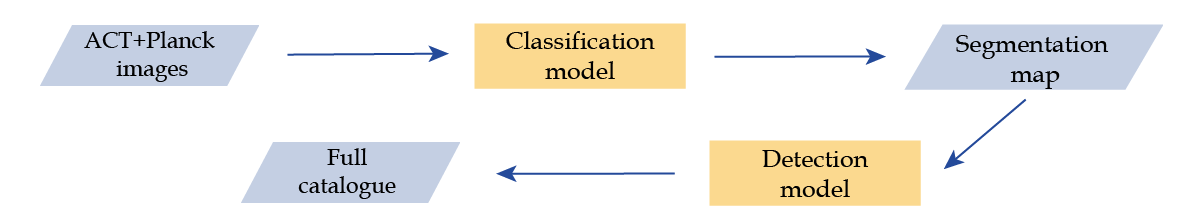}
    \caption{Scheme showing main steps of building the catalogue of potential SZ sources}
    \label{fig:scheme}
\end{figure*}

\thispagestyle{empty}
Fig.~\ref{fig:scheme} illustrates the sequence of pipeline steps in constructing a catalogue of potential SZ sources. We convert the full ACT+Planck intensity map into a SZ signal segmentation map by using a Deep Learning classification model (see Section~\ref{sec:DL_class}). Then, the detection procedure (described in Section~\ref{sec:detection}) is applied to the SZ signal segmentation map, resulting in a catalogue of SZ sources candidates. Below, we describe the SZ signal segmentation and source detection procedures in detail. 

\subsection{Deep learning SZ signal segmentation model}
\label{sec:DL_class}
To recognise SZ signal on ACT+Planck intensity maps, one could follow the approach already presented in \cite{SZcat,Bonjean_2020}, i.e. to train a classical U-Net segmentation architecture on the microwave data. To do so, one needs to define a ground truth segmentation mask. For example, \cite{SZcat,Bonjean_2020} for each source in the training sample of clusters drew circles at the cluster positions with the average Planck PSF FWHM size. This approach implies that a galaxy cluster is approximated as a circle with the Planck PSF radius, thus some information about the real shape and size of galaxy clusters in the training set is lost.  

In this work, we use a different approach to SZ signal deep learning segmentation. We train a DL classification model on snapshot images centred on ACT DR5 galaxy clusters (the positive class), snapshot images centred on radio sources (mainly, dusty star-forming galaxies and
radio-loud AGNs) and snapshot images taken in random directions of the sky (the negative class). Then, we apply our classifier to each pixel of ACT+Plank intensity maps and thus obtain the SZ signal segmentation map in the ACT footprint. This approach has a potential benefit of retaining information about clusters shapes in the deep learning model. 

Below we explain a construction of SZ signal classification model.

\subsubsection{Classification model architecture}
\label{sec:acrh}

\begin{table}
\centering
\begin{tabular}{c c c } 
 \hline
  & Layer & Output map size \\ [0.0ex] 
 \hline\hline
 1 & Input & $32 \times 32 \times 3$ \\ 
 2 & $5 \times 5$ convolution  +ReLU& $32 \times 32 \times 16$ \\
 3 & $3 \times 3$ convolution +ReLU   & $32 \times 32 \times 16$ \\ 
 4 & $2 \times 2$ MaxPooling  & $16 \times 16 \times 16$ \\
 \hline
 5 & $3 \times 3$ convolution  +ReLU& $16 \times 16 \times 32$ \\
 6 & $3 \times 3$ convolution +ReLU   & $16 \times 16 \times 32$ \\ 
 7 & $2 \times 2$ MaxPooling  & $8 \times 8 \times 32$ \\
 \hline
 8 & $3 \times 3$ convolution +ReLU & $8 \times 8 \times 64$ \\
 9 & $3 \times 3$ convolution +ReLU   & $8 \times 8 \times 64$ \\ 
 10 & $2 \times 2$ MaxPooling  & $4 \times 4 \times 64$ \\
 \hline
 11 & $3 \times 3$ convolution +ReLU & $4 \times 4 \times 128$ \\
 12 & $3 \times 3$ convolution +ReLU  & $4 \times 4 \times 128$ \\ 
 13 & $2 \times 2$ MaxPooling  & $2 \times 2 \times 128$ \\
 \hline
 14 & Flatten  & $512$ \\
 15 & Linear+ReLU     & $256$ \\
 16 & Linear+ReLU     & $128$ \\
 17 & Linear+ReLU     & $32$ \\
 18 & Linear+Sigmoid  & $1$ \\
 \hline
\end{tabular}
\caption{Our model architecture. All layers, except the last one, use rectified linear units, and all the convolutional and linear layers use the batch normalisation.}
\label{table:arch}
\end{table}

We aim to construct a classification algorithm that analyses a segment of a microwave ACT+Planck map and yields a probability of detecting a cluster at a centre of a segment. To do so, we use a convenient CNN+MLP architecture, the adopted version of VGG \citep{VGG}). 

Table~\ref{table:arch} summarises the architecture of the artificial neural network model. The CNN (convolution neural network) part contains 4 identical blocks. Each convolutional block (except for the first one, in which the first convolution has a kernel of size 5) has two convolution kernels of 3x3 pixels size (padding=1, stride = 1), the rectified linear unit activation function (ReLU$(x) = \max(0, x)$; \citet{ReLU}) and the MaxPooling sub-sampling unit. The second part contains a Multi-Layer Perceptron (MLP) with three linear layers, each with a ReLU activation function. We use the batch normalisation \citep{batchnorm} in CNN and MLP layers and DropOut with $p=0.5$ \citep{Dropout} in each MLP layer for the network regularisation. We use the linear layer with the standard Logistic Sigmoid function (see e.g. \cite{2021arXiv210914545D}) to obtain a probability of SZ signal as the network output. Total number of trained parameters equals to 462,865. 

\subsubsection{Model training and validation}
\label{subsec:train_test_sample}
To create the model input, the ACT+Planck map is divided into two distinct regions: training and test sets. The training set is used to learn the optimal model parameters, while the test set is needed to evaluate a final model. We also include in the knowledge base (train and test samples) fields in random directions on the sky, which do not overlap neither with clusters from ACT DR5 nor with radio sources (within a 10-minute radius).

For the training set, we select the western region of the sky in equatorial coordinates ($\alpha>180^\circ$). This region encompasses a total of 2,449 galaxy clusters from ACT DR5 ($\simeq $ 35.5 \% of the train data set) and 1,226 non-cluster radio sources ($\simeq $ 17.8 \%). Random fields comprise approximately  46.7\% of the train sample. 

To prepare our dataset, we cut the ACT+Planck maps into 16 arcmin x 16 arcmin squares (32 x 32 pixels). To improve the model accuracy and to avoid overfitting, we apply image preprocessing and augmentation techniques to our input images. In particular, we normalise all the images in the following way:
\begin{align}
    x = \frac{x}{max(||x||_2, 0)}, 
\end{align}
where $||x||_2$ is the Euclidean norm over rows. Further image transformations include standard augmentation techniques such as flipping and random perspective.

To quantify the model performance, we use a binary cross-entropy loss function $\ell(y, f_\theta)$: 
\begin{align}
 \ell(y, f_\theta) = -y\log(f_\theta) - (1-y)\log(1-f_\theta)
\end{align}
where $y$ is the actual class ($y=0$ for non-clusters and $y=1$ for clusters), and $f_\theta \in [0, 1]$ is the predicted label, which can be interpreted as the probability of detecting a cluster in a given field. The loss function compares predicted probabilities to the actual class output and calculates a penalty score based on the distance from the expected value. This loss function is a way to measure the accuracy of a deep learning algorithm in predicting expected outcomes and is useful in optimising the model and increasing its accuracy. 

For iterative updating  the network parameters $\theta$, we use the Adam optimisation method \citep{Adam} (with default parameters except weights\_decay $= 0.001$). All hyper-parameters, such as kernel size, padding, stride, etc., are selected manually by training different models to achieve the best performance.

\subsubsection{Classification model test}
\label{subsec:test_sample}
The test data set includes 1,745 galaxy clusters ($\simeq $ 30.7 \% of the test data set), 519 point sources ($\simeq $ 9.1 \%), and 3,422 random fields ($\simeq$ 60.2\%) in the eastern part of the sky ($\alpha\le{}180^\circ$). 

\begin{figure}
\centering\includegraphics[width=0.48\textwidth]{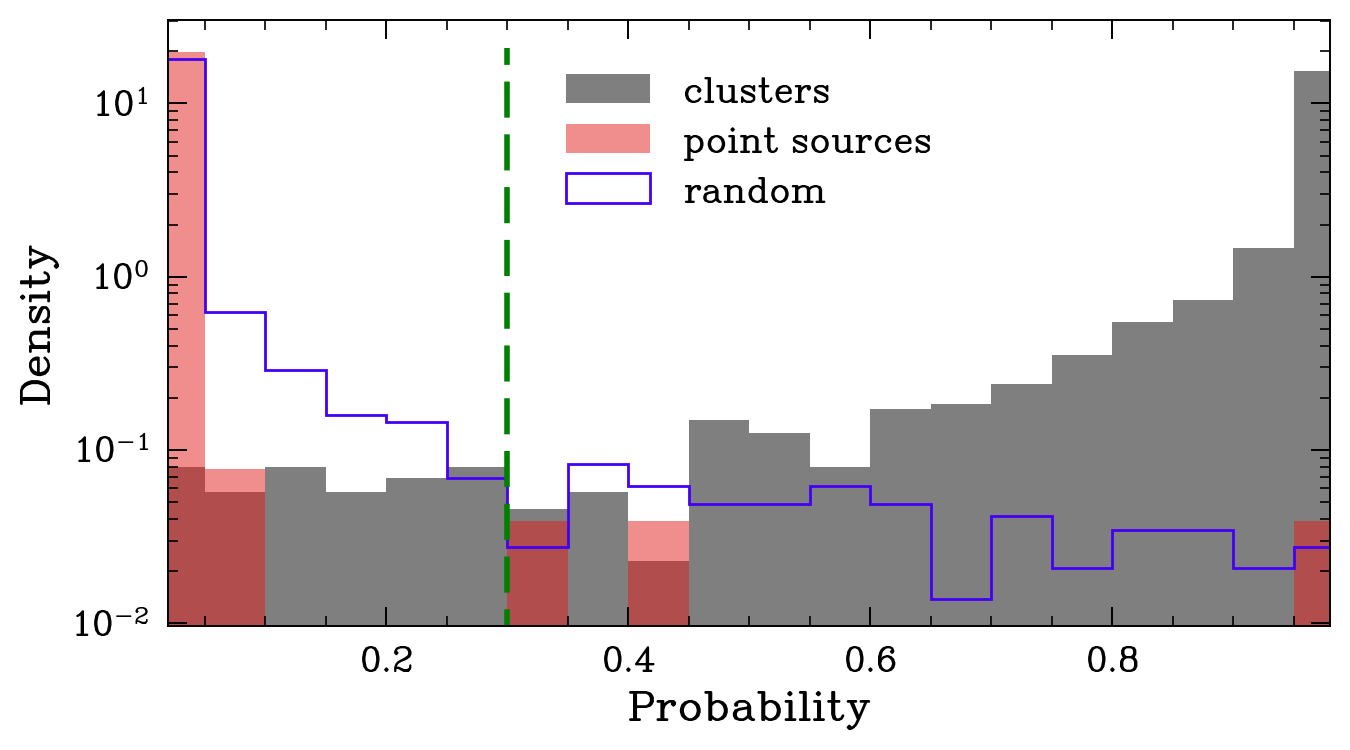}
\caption{Distribution of probabilities predicted by our model for the test dataset: galaxy clusters are shown in grey, points sources are in red, and random fields are shown with purple curve. Each histogram is normalised by the total amount of objects of a given type. The green vertical line marks the probability threshold below which most of objects are actually non-clusters. We use this threshold in Section~\ref{sec:detection} to distinguish between non-clusters and cluster candidates.}   
\label{fig:curves}
\end{figure}

To demonstrate the model behaviour, we show a distribution of predicted probabilities of detecting a cluster in a given patch of the sky for the test sample on Figure~\ref{fig:curves} (we consider only the area with the galactic latitude $|b| > 15^\circ$ to avoid the Galactic plane). The test dataset includes clusters, point sources and random fields (see Section~\ref{subsec:train_test_sample}). From Figure~\ref{fig:curves} we see that for most of the clusters (in grey) from the test dataset, the model correctly assigns high probabilities (note that the vertical axis is shown in log-scale), while for most of the point sources (in red) and random fields (in blue) predicted probabilities are low. Approximately $\sim $0.06\% of point sources, the model misclassifies as galaxy clusters. The vertical line marks the probability threshold, which we further use to form a criterion for distinguishing cluster candidates from non-clusters. Most of the objects with probabilities $<0.3$ are non-clusters. Although, $\sim $2\% of genuine clusters are not recognised as clusters by our algorithm (i.e. the model predicts $p<0.3$). There is a $\sim $3 \% misclassification ($p>0.3$) in random fields in test, i.e. some random fields are recognised as clusters.

In general, in data-driven deep learning-based modelling, data quality may affect classification performance \citep{NoisyLabels}. Let us discuss possible sources of erroneous labels in the training data set  and how such labels may impact the overall performance of our model. False labels in the ACT DR5 catalogue are not expected since all clusters there are optically confirmed. Erroneous labels in the negative class (which contain AGN, DSFG, and random (empty) fields) might be present, because some of the snapshots (more precisely, the central pixel of the snapshots) might contain a serendipitous cluster. We roughly estimate a probability of detecting a cluster in a given pixel by summing up the area\footnote{assuming that a cluster extends up to its $R_{500}$} occupied by  clusters from ACT DR5  and dividing it by the total area covered by the catalogue (see~Table~\ref{table:summary}). Then we multiply the resulting probability by the number of snapshots in the negative class of the train data set. So we conclude that  $\sim $ 10 clusters might be present in the non-cluster class, or, in other words, there are $\sim $ 10 erroneous labels (less than 0.1 \%) in our data-set. Taking into account that the loss function is calculated by averaging values obtained in batches of size 64 objects, the influence of a few erroneous labels is not significant.

\begin{figure*}
\centering\includegraphics[width=0.66\textwidth]{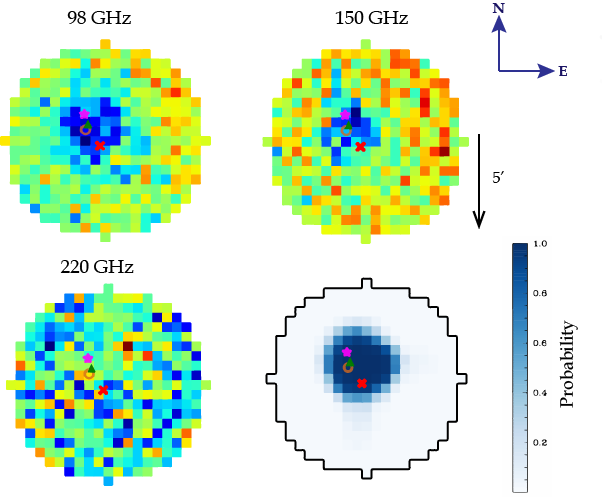}
\caption{An example of a cluster detection in the direction of SZ\_G098.31-41.19 (marked as the red cross) from the SZcat catalogue. 5-arcmin fragments of the ACT+Planck maps at different frequencies are shown in the upper row and in the bottom left panel. The pixel size is 0.5 arcmin. The probability map is presented at the bottom right corner. For comparison, we also show positions of clusters from PSZ2 (magenta star) and ACT DR5 (brown circle) catalogues which are within the cross-match radius, i.e. 5 arcmin circle. The pixel with maximum probability is marked as a green triangle.} 
\label{fig:example}
\end{figure*}

\subsubsection{Segmentation map}
\label{subsec:segmentation}
We apply our classifier to each pixel of ACT+Planck intensity maps and obtain a segmentation map. Each pixel in the segmentation map denotes a probability of a presence of the SZ signal in the direction of pixel centre. 

An example of a patch of the segmentation map containing a massive galaxy cluster is shown in Fig.~\ref{fig:example}. We choose the direction of SZ\_G098.31-41.19 (marked as the red cross) from the SZcat catalogue. The top row and the bottom left panel show three fragments of the ACT+Planck input maps at 98, 150, and 220 GHz. The probability map for this area is shown at the bottom right. It contains only one connected group within a chosen window, and this group can be associated with the galaxy cluster PSZ2 G098.30-41.15 from the PSZ2 catalogue (marked with a magenta star) and ACT-CL J2334.3+1759 from ACT DR5 (its centre is shown with a brown circle). It's a massive cluster with $M_{500} \sim (7-8) 10^{14}$ $M_{\odot}$ \citep{Hilton_2021} located at $z=0.436$.

\subsection{Detection model}
\label{sec:detection}
Since galaxy clusters are extended objects, in the segmentation map we look for connected groups of pixels with probabilities above a certain threshold. To do so, we use the scikit-image \citep{scikit-image} library developed for scientific image analysis in Python. We analyse manually a sample of probability maps for known clusters (i.e. patches of the segmentation map which contain galaxy clusters) and random fields on the sky (with no clusters) and derived the optimal value of the threshold segmentation $p_{seg}=0.3$ (see Fig.~\ref{fig:curves}).  For comparison, \cite{Bonjean_2020} used a threshold $p_{seg}=0.1$ to select sources in the SZ signal segmentation map created from Planck HFI imaging data.

Additionally, we remove the complex Galactic plane from consideration and keep only objects with the galactic latitude $|b| > 15^{\circ}$. The resulting catalogue contains more than $1$ million potential SZ sources candidates with $p>0.3$ in the ACT footprint. 

\section{The ComPACT catalogue}
\label{sec:compact}
\thispagestyle{empty}
Here, we apply our method to the extended Planck cluster candidates sample SZcat \citep{SZcat}, having a low purity, in order to get a medium-sized cluster catalogue, ComPACT, dominated by SZ objects seen in Planck, and having a fairly high cluster purity. By its construction, our resulting catalogue does not contain any new detections relative to SZcat, but a certain fraction of clusters is barely distinguishable from the noise in the Planck maps but seen in the ACT data (see Appendix~\ref{app:metrics}). That's why we consider ComPACT as a combined ACT+Planck cluster catalogue.
\begin{figure*}
\centering
\includegraphics[width=1\textwidth]{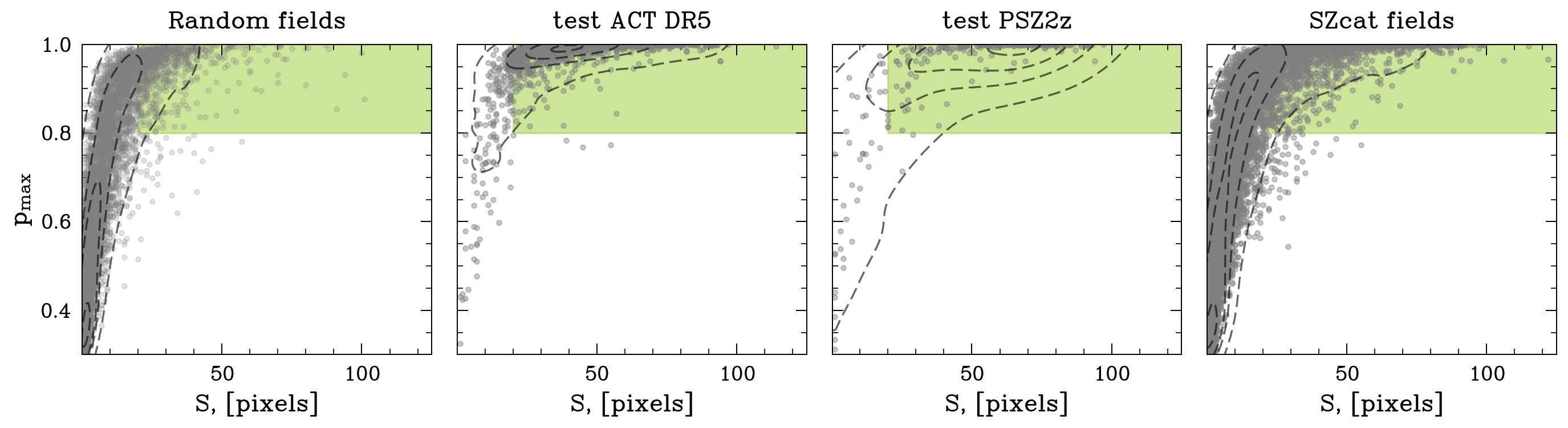}
\caption{Group area $S$ as a function of the  maximum probability within a group $p_{max}$. From left to right: random fields (in the ACT footprint), groups associated with optically confirmed galaxy clusters from the test ACT DR5 data set and from the test PSZ2z subsample (in the ACT footprint), and SZcat directions (in the ACT footprint). With green colour we show a region where detected groups of pixels are classified as clusters: $S>20$ and $p_{max}>0.8$. We see that with this simple criterion in random fields almost all detected sources are classified as non-clusters (as expected), with only $\sim 10$\% of groups falling into a 'cluster region'. For ACT DR5 and PSZ2 optically confirmed clusters, we correctly classify $\sim 80$\% of clusters}
\label{fig:SP}
\end{figure*}

\begin{figure*}
\centering
\includegraphics[width=0.9\textwidth]{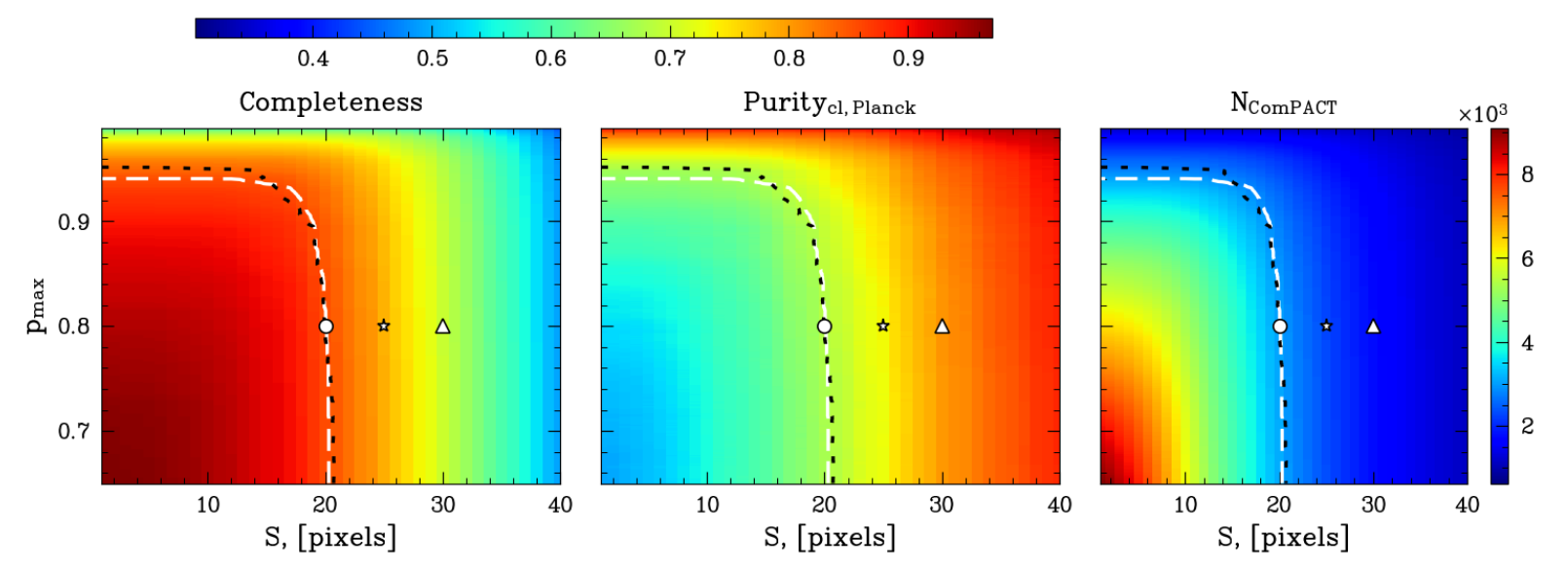}
\caption{An illustration how the resulting catalogue characteristics depend on our criterion (in terms of $S$ and $p_{max}$) of detecting a cluster. Left panel shows the catalogue completeness with respect to ACT+PSZ2z clusters, the middle panel presents the catalogue purity with respect to Planck clusters, while the right panel demonstrates the  number of cluster candidates in the resulting catalogue.  White dot shows \emph{'Priority'} = 3 cluster candidates  ($S \ge 20$ \& $p_{max} \ge 0.8$), white star corresponds to  \emph{'Priority'} = 2 objects ($S \ge 25$ \& $p_{max} \ge 0.8$), white triangle marks \emph{'Priority'} = 1 subsample (i.e. the most reliable clusters with $S \ge 30$ \& $p_{max} \ge 0.8$). In each panel, the dashed white line shows the contour with recall $C = 0.86$, while the dotted black line shows the contour with $Purity_{cl,Planck} = 0.67$. We see that as the completeness and $N_{ComPACT}$ decrease, the $Purity_{cl,Planck}$ in the sample increases.}
\label{fig:params}
\end{figure*}

For each direction in the SZcat catalogue, we consider the nearest detected connected group of pixels in a window with $R_{match} = 5$ arcmin. For each object we calculate (i) the group area $S$ (the total number of pixels in a given connected group), (ii) the maximum probability $p_{max}$ in an object, and (iii) the distance $R$ from the object centre (the pixel with $p_{max}$) to the input direction (see Appendix~\ref{sec:appC}). Below we analyse different samples to form a criterion for distinguishing clusters from non-clusters among the connected groups (Figures~ \ref{fig:SP} and \ref{fig:quan} in Appendix~\ref{sec:appC}). 

Let us now illustrate how a distribution of objects looks like on the $p_{max}$--$S$ plane for random directions\footnote{but within the ACT field} on the sky (see the left panel of Fig.~\ref{fig:SP}). Random fields are unlikely to contain galaxy clusters. We see that non-clusters tend to have low areas and lower values of $p_{max}$ compared to clusters. In central panels of Fig.~\ref{fig:SP}, we show distributions of clusters detected by our model along the input directions from the ACT DR5 test subsample (see Section~\ref{sec:dataset}) and from test PSZ2z (PSZ2z is the PSZ2 subsample of clusters with spectroscopic redshifts). All those objects are optically confirmed clusters. We see that real clusters tend to have large areas and large values of $p_{max}$. Right panel of Fig.~\ref{fig:SP} presents the $p_{max}$--$S$ distribution of the nearest objects detected by our model in the 5 arcmin window along the SZcat directions. We detect a large number of sources with high area and high values of $p_{max}$, which are likely to be genuine clusters. 

In order to select optimal thresholds to separate clusters from non-clusters, we analyse the behaviour of the following three characteristics whose dependence on the $S-p_{max}$ threshold grid is shown in Figure~\ref{fig:params}:
\begin{enumerate}
    \item Completeness (see eq.~\ref{eq:completness}) on the left panel. We estimate completeness with respect to optically confirmed clusters from PSZ2z + ACT DR5 catalogues. We associate clusters from the PSZ2z+ACT sample in the test area with the nearest object from our full catalogue of SZ candidates (see Section \ref{sec:detection});
    \item Purity with respect to Planck clusters (see eq.~\ref{eq:purity_planck}) on the central panel. This parameter is estimated from the number of random fields $\hat{N}_{field}$ in which our model detects a cluster (see Table~\ref{tab:sth} for typical $\hat{N}_{field}$ values for different area thresholds). We consider $\hat{N}_{field}$ as the number of false positive detections with respect to the Planck clusters;
    \item The number of clusters $N_{ComPACT}$ in a resulting catalogue  on the right panel. 
\end{enumerate}

In Figure~\ref{fig:params}, we show how the above mentioned metrics change with respect to different $S$ and $p_{max}$ thresholds. As the area $S$ and $p_{max}$ decrease, the number of objects in a catalogue and the catalogue completeness  increase, but the percentage of Planck clusters, "recovered" in the ACT data by our model, decreases. Therefore, it is important to choose thresholds that ensure high completeness and purity of the resulting catalogue. To strike a balance between the two, we choose the thresholds $S = 20$ and $p_{max} = 0.8$, which provide a compromise between the completeness and $Purity_{cl,Planck}$. The resulting sample of objects with $S \ge 20$ and $p_{max} \ge 0.8$ we call the ComPACT catalogue, and consider as one of the main deliveries of this work. Contours are plotted with $C = 0.86$ (the dashed white line) and $Purity_{cl,Planck} = 0.68$ (the dotted black lines). Choosing a higher threshold for $p_{max}$ results in a loss of purity, so we select thresholds where the white and black  contours intersect. Additionally, we  create  a column  \emph{'Priority'} which indicates the reliability of clusters:  
\begin{itemize}
    \item \emph{'Priority'} = 1 is given to cluster candidates with  $S \ge 30$ and $p_{max} \ge 0.8$ (marked with the triangle in Fig.~\ref{fig:params});
    \item \emph{'Priority'} = 2 is assigned to objects with $S \ge 25$ and $p_{max} \ge 0.8$ (the star in Fig.~\ref{fig:params});
    \item \emph{'Priority'} = 3 is for objects with $S \ge 20$ and $p_{max} \ge 0.8$ (the circle in Fig.~\ref{fig:params}).
\end{itemize}  
By definition, \emph{'Priority'} = 1 subsample has the highest purity among subsamples with different priorities.

\begin{figure}
    \centering
    \includegraphics[width=1\linewidth]{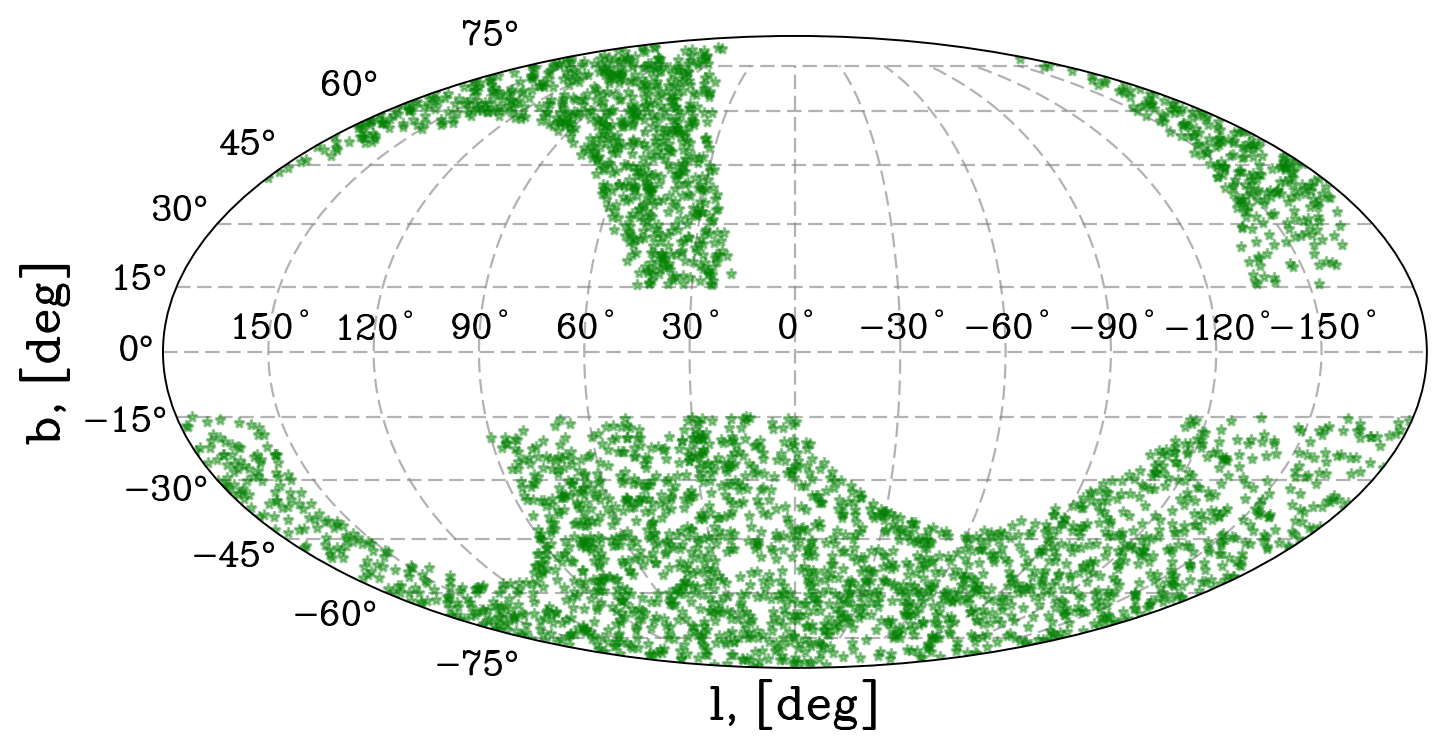}
    \caption{Distribution of ComPACT objects on the sky (in galactic coordinates)}
    \label{fig:map}
\end{figure}

\begin{table*}
\centering
\begin{tabular}{c c c c c} 
 \hline
 Cluster catalogue & Number of objects & Sky fraction & Instrument & Ref\\ [0.0ex] 
 \hline\hline
 SZcat     & 30,917 & 0.95  & Planck & \cite{SZcat}\\ 
 PSZ2      & 1,653  & 0.74  & Planck & \cite{psz2} \\ 
 ACT DR5   & 4,195  & 0.35  & ACT    & \cite{Hilton_2021}\\ 
 SPT-SZ    & 677    & 0.07  & SPT    & \cite{SPT-SZ}\\ 
 \hline
 ComPRASS  & 2,323  & 0.79  & Planck, ROSAT & \cite{ComPRASS}\\ 
 PSZSPT    & 419    & 0.06  & SPT, Planck   & \cite{PSZSPT}\\
 \hline
 MCXC      & 1,743  & 0.67  & ROSAT, EXOSAT & \cite{mcxc}\\ \hline
ComPACT    & 2,962  & 0.37  & ACT+Planck    & this paper \\
 \hline
\end{tabular}
\caption{External galaxy cluster samples in comparison with new ComPACT cluster catalogue presented in this paper. We estimate the sky fraction by summing all healpix cells (with nside=$2^3$), which contain at least one galaxy cluster from the considered catalogue.}
\label{table:summary}
\end{table*}

In Fig.~\ref{fig:map} we show a distribution of ComPACT clusters on the sky (in galactic coordinates).

\subsection{Cross-correlation of ComPACT with external cluster catalogues}
\label{subsec:inters}
\begin{figure*}
\centering
\includegraphics[width=1\textwidth]{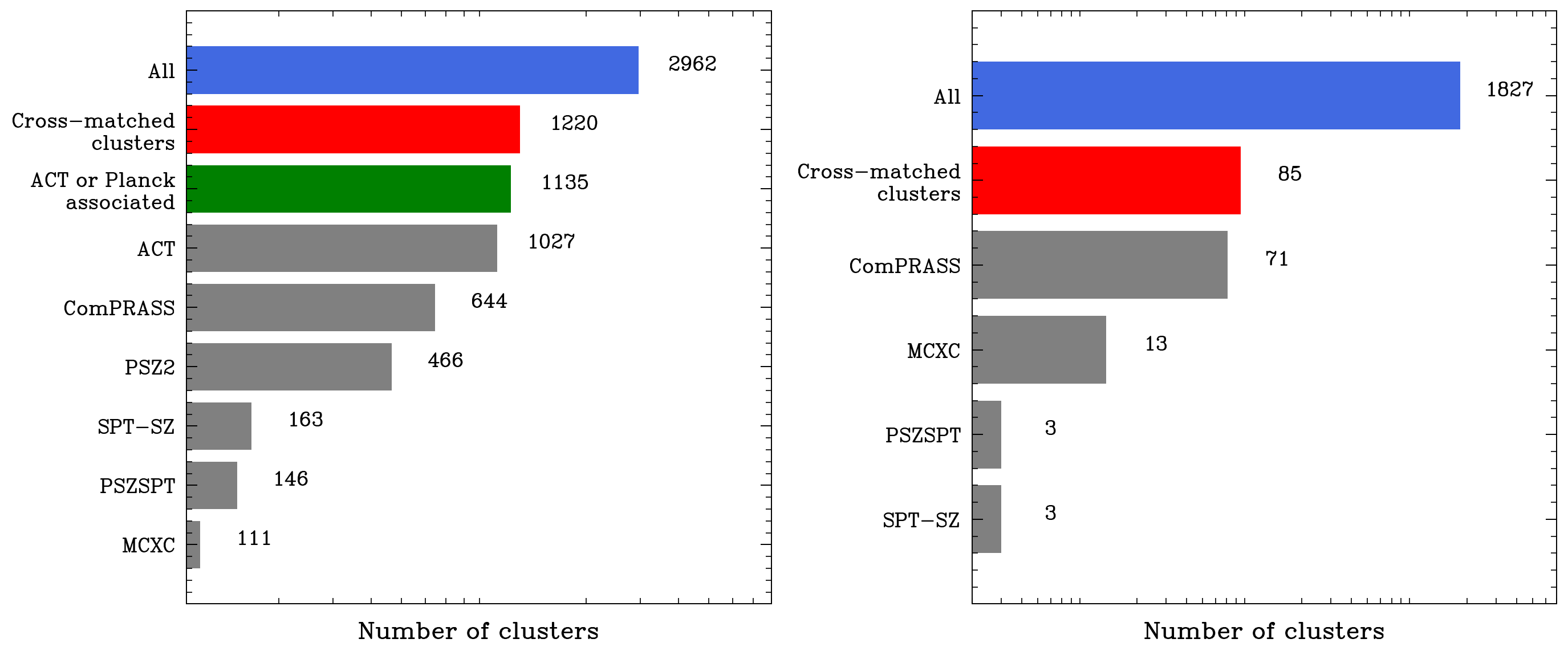}
\caption{Number of cross-matches of ComPACT cluster candidates with other catalogues. The left panel shows all candidates, the right panel shows only ComPACT candidates that have no matches with ACT DR5 or PSZ2 catalogues. In each panel, the first line shows the total number of ComPACT objects in a considered sample/subsample (in blue). The second line illustrates the total number of matches across explored catalogues. The remaining lines represent the number of matches with a given catalogue. }
\label{fig:inters}
\end{figure*}

We cross-correlate the derived ComPACT sample with the following X-ray and SZ external cluster catalogues: MCXC, ACT DR5, PSZ2, SPT-SZ, ComPRASS, and PSZSPT. The cross-match radius is set to 5 arcmin. This choice is motivated by the analysis of average number density radial profiles (for different cluster catalogues) in the neighbourhood of ComPACT objects (see Figure~\ref{fig:r}). Table~\ref{table:summary} gives the summary information about external cluster catalogues in comparison with ComPACT. 

In Fig.~\ref{fig:inters} (left panel), we show numbers of ComPACT clusters associated with different external cluster catalogues. ComPACT contains $2,962$ cluster candidates (see Table~\ref{tab:sth} for numbers related to different \emph{'Priority'} subsamples). In total, $1,220$ ComPACT objects are associated with sources from external cluster catalogues, among which $1,135$ clusters are found in Planck and/or ACT cluster catalogues. From ACT DR5, we identify $1,027$ matches, cross-correlation with the PSZ2 catalogue gives $466$ clusters.  

In Fig.~\ref{fig:inters} (right panel), we show numbers of ComPACT clusters that are not found in PSZ2/ACT cluster catalogues (also see row $N_{new}$ in Table~\ref{tab:sth}) and are associated with objects from other cluster samples. $1,827$ ComPACT objects are new with respect to the existing ACT DR5 or PSZ2 cluster samples. One can see, that 85 objects are found in MCXC, SPT-SZ, ComPRASS, PSZSPT catalogues. We identify 3 clusters in SPT-SZ, 3 clusters in PSZSPT, 13 --- in MCXC and 71 --- in ComPRASS. All ComPACT objects identified as X-ray galaxy clusters with luminosity (in the energy band 0.5-2 keV) exceeding $7.5 \cdot 10^{44}\:erg/s$ have $p_{max}\sim1$. 

\begin{table}
    \centering
    \begin{tabular}{cccc}
    \hline
       \emph{'Priority'}  & 3: $S > 20$ & 2: $S > 25$ & 1:$S > 30$\\
         \hline\hline
       $N_{ComPACT}$ & 2,962 & 2,218 & 1,720\\
       $N_{PSZ2}$ & 466 & 447 & 422\\
       $N_{new}$ & 1,827 & 1,139 & 721\\ 
       \hline
       $\hat{N}_{field}$ & 946 & 596 & 337\\
       \hline
       $Purity_{cl, min}$ & 0.74 & 0.78 & 0.84\\
       $N_{new, cl}$ & 977 & 589 & 403\\
       \hline
       $Purity_{cl, Planck}$ & 0.68 & 0.73 & 0.8\\
       $N_{new, cl}$ & 794 & 471 & 331\\
       \hline
       $Purity_{cl, high}$ & 0.94 & 0.95 & 0.96\\
       $N_{new, cl}$ & 1,564 & 959 & 606\\
       \hline
       $C$ & 0.86 & 0.78 & 0.7\\
       $C_{ACT}$ & 0.88 & 0.81 & 0.71\\
       $C_{PSZ2z}$ & 0.7 & 0.66 & 0.61\\
       
       \hline
    \end{tabular}
    \caption{The table displays catalogue statistics based on \emph{'Priority'}. The first three rows show the number of objects in the catalogue: the total number, the number associated with PSZ2z, and the number of new objects relative to ACT/PSZ. The following row shows the number of random fields that fall inside the thresholds. The next three blocks display the purity estimates and the resulting estimates of the number of new real clusters. The final block pertains to estimates of completeness: the first row pertains to ACT+PSZ2z, the second to ACT, and the third to PSZ2z.}
    \label{tab:sth}
\end{table}

\begin{figure*}
    \centering
    \begin{minipage}[!h]{1\linewidth} 
     \includegraphics[width=1\textwidth]{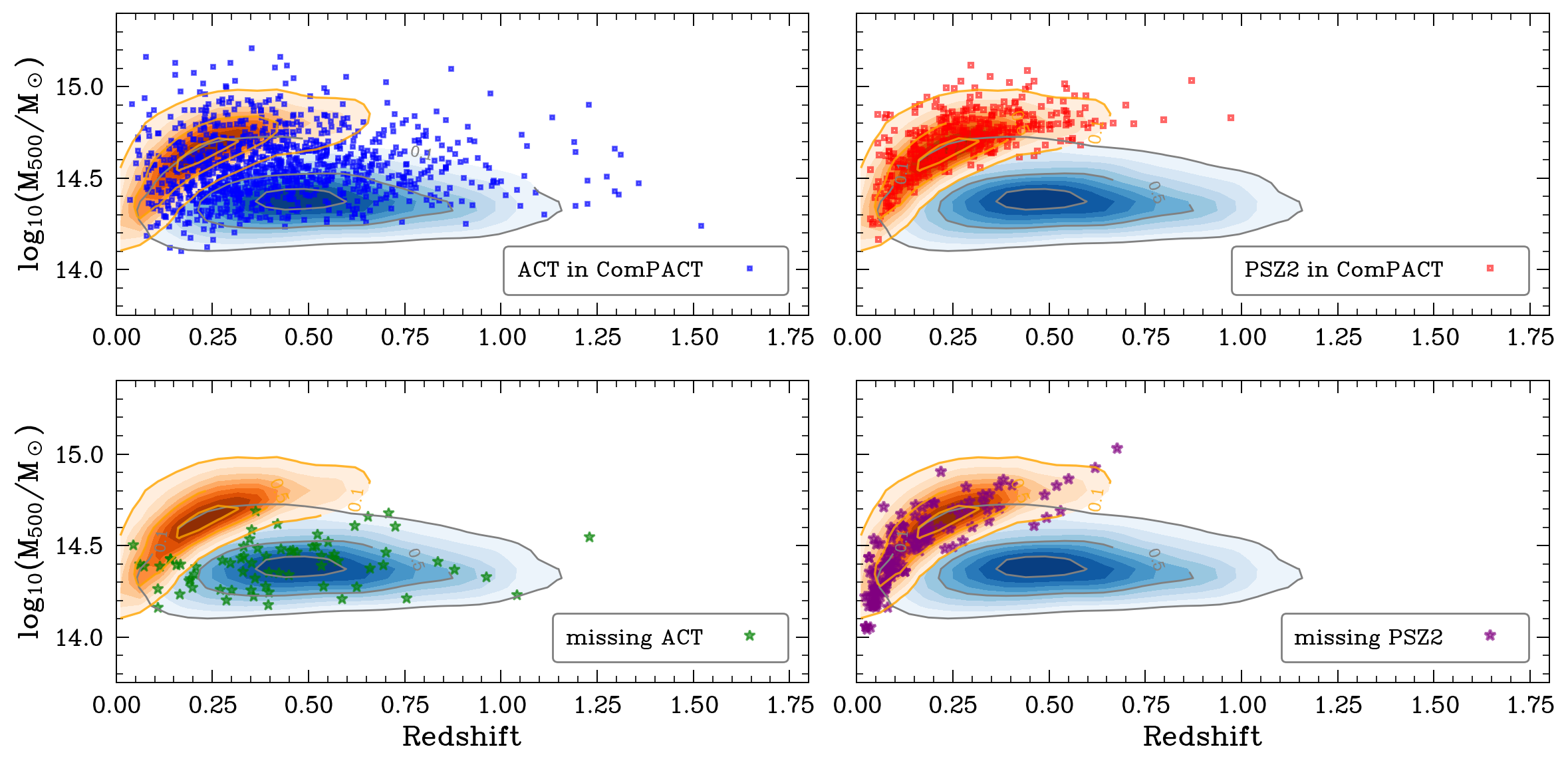}
    \caption{ComPACT cluster candidates cross-matched with ACT DR5 (1,027 objects; blue dots in the upper left panel) and PSZ2z (399 objects; orange dots in the upper right panel). On the bottom panel we present clusters from ACT DR5 (green stars) and PSZ2z (purple stars) that have SZcat candidates within a 5 arcmin radius, but are not presented in ComPACT. Blue contours show density of SZ sources in ACT DR5 on the mass-redshift plane, while orange contours represent density of PSZ2z sources. From comparing ComPACT candidates matched with ACT DR5 and density contours of ACT DR5 sources, it appears that we miss low-mass clusters especially at intermediate and high redshifts. It might be due to our cluster detection criteria and because ComPACT is based on SZcat, which in turn is constructed from Planck maps. }
    \label{fig:SZ_ap}
    \end{minipage}
    \vfill
    \begin{minipage}[!h]{0.49\linewidth}
    \center{\includegraphics[width=1\linewidth]{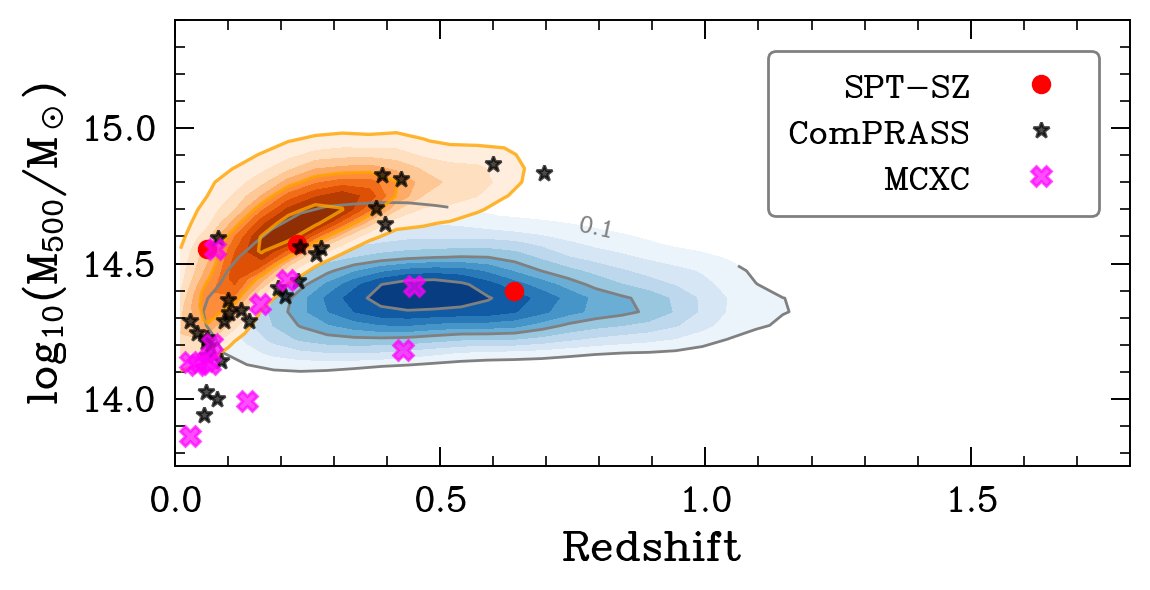}}
    \caption{Mass-redshift dependence for ComPACT cluster candidates which do not have matches with ACT DR5 or PSZ2 but do have with MCXC (magenta crosses), ComPRASS (black stars), and SPT-SZ (red circles). The orange density contours show density of PSZ2z sources, the blue one --- ACT DR5 clusters.}
    \label{fig:new_Mz}
    \end{minipage}
\end{figure*}

\subsection{Catalogue metrics}
\label{sec:metrics}

The key characteristics of a cluster catalogue are its purity and completeness metrics. In Table~\ref{tab:sth} we show how the key characteristics of the catalogue change with the \emph{'Priority'} value: $N_{ComPACT}$ --- the number of objects in a catalogue, $N_{PSZ2}$ --- the number of ComPACT clusters associated with PSZ2, and $N_{new}$ --- objects that are not found in ACT/PSZ2 samples. One can see that all these numbers decrease as the \emph{'Priority'} increases. In the catalogue with \emph{'Priority'} = 3 there are 2,962 candidates, and the cleanest sample contains 1,720 objects.  In Table~\ref{tab:sth} we also provide $\hat{N}_{field}$ --- the number of detections with chosen thresholds ($p_{max}$, $S$) in the random field directions. Then, we introduce the following estimates of purity: 
\begin{enumerate}
    \item $Purity_{cl, Planck}$ (see eq.~\ref{eq:purity_planck}) reflects  the percentage of the Planck clusters found in the catalogue;
    \item $Purity_{cl, min}$ (see eq.~\ref{eq:purity_min}). The lower purity is estimated by using SZcat purity estimate of 21.7 \%. We expect a higher purity, as we find more probable directions using our method;
    \item $Purity_{cl, max}$ (see eq.~\ref{eq:purity_high}). In this evaluation, we used information about false detections that can be estimated by counting $\hat{N}_{field}$ and $p_{max}$, which is good corrected on the test sample.
\end{enumerate}

In Table~\ref{tab:sth} we show the estimate of purity and $N_{new, cl}$ --- number of new clusters that are expected to be real with respect to ACT/PSZ2. $\hat{N}_{field}$ decreases from 946 to 337, so number of false detections decreases due to area threshold. We see that all purities increase from 74/94 \% to 84/96 \% as the threshold of S increases. 

We estimate the ComPACT completeness with respect to PSZ2z (the subsample of PSZ2 objects with optical identification and cluster spectroscopic redshift measurements) and ACT DR5 cluster catalogues. We associate clusters from PSZ2z or ACT samples in the test area with the nearest object from our full catalogue of SZ candidates (see Section \ref{sec:detection}). All cluster associations are shown in the central panels of Figures~\ref{fig:quan} and \ref{fig:SP}. Then, we estimate the ComPACT catalogue completeness as in eq.~\ref{eq:completness}. 

In Table~\ref{tab:sth} we show completeness for three \emph{'Priority'} tags for the ACT DR5 and PSZ2z cluster samples. For ACT DR5 + PSZ2z sample we get $C \approx 0.7 -0.86$, then for ACT DR5 separately $C_{ACT}\approx0.71 - 0.88$ and PSZ2z --- $C_{PSZ2z}\approx0.61 - 0.7$. Some percentage of true clusters are not assigned to clusters by our (detection) model due to their low area, i.e. these objects have $S<20$ (\emph{'Priority'} = 3), $S<25$ (\emph{'Priority'} = 2) and $S<30$ (\emph{'Priority'} = 1). The decrease in the completeness of Planck cluster detection in ComPACT seems to be due to: (i) the way of combining ACT and Planck data in the joint intensity maps \citep{Naess_2020} and (ii) the wrong association between the connected group and SZcat for some ComPACT clusters.

Another variant of completeness loss is the appearance of AGN in the cluster centre. To check the influence we cross-correlate  ACT DR5 test sample with a QSO from SDSS DR17 in 1.4 arcmin (a diffraction-limited resolution of ACT in 150 GHz). We find 93 clusters with quasars in the centre. We then construct a cumulative function for the sample with and without QSOs, to which we apply the KS test. We check whether the samples belong to the same statistical population and obtain $p_{value} = 7.6 \cdot 10^{-6}$. This shows that the AGN in the centre of the cluster do not affect the predictions of the model.

\subsection{Mass-redshift}
\label{subsec:mz}
In Figure~\ref{fig:SZ_ap}, we illustrate the mass-redshift dependence for clusters in ACT DR5 and PSZ2z catalogues\footnote{Here, we do not aim to compare masses of clusters obtained in different catalogues. The goal is rather to illustrate typical values for detected clusters. However, \cite{Hilton_2021} (see their Figure 21) demonstrated that  masses from PSZ2z and ACT DR5 (the M500cUncorr column) on average agree for a subsample of clusters present in both catalogues}. The blue and orange contours represent the distributions of the full ACT DR5 and PSZ2z catalogues, respectively. For ACT DR5 cross-correlated with SZcat, we `recover' 1,027 clusters (shown as blue squares in the upper left panel), and 70 ACT clusters\footnote{with SZcat candidate in $R_{match} = 5$ arcmin} are missing in ComPACT (shown as green stars in the bottom left panel). We miss low-mass cluster candidates at intermediate and high redshifts due to (i) detection limitations of the SZcat catalogue, and (ii) our area criterion (see the second panel of Figure~\ref{fig:SP}).

The mass-redshift distribution for PSZ2z matches (for candidates with measured masses and redshifts, 399 objects) in ComPACT is plotted as red squares in the upper right panel of Fig.~\ref{fig:SZ_ap}. PSZ2z clusters that are not present in ComPACT are shown as purple stars (154 objects) in the bottom right panel of Fig.~\ref{fig:SZ_ap}.  

Let us now discuss cluster candidates that do not have any matches in ACT DR5 or PSZ2 catalogues. In Figure~\ref{fig:new_Mz}, we plot matches (with available in the literature masses and redshifts) with SPT-SZ (red dots), ComPRASS (black stars), and MCXC (magenta crosses). Most of them lie at $z < 0.8$. Compared to the ACT DR5 catalogue, ComPACT is more complete at lower redshifts. We expect to find objects with low mass at low redshifts and clusters on the border of PSZ2z density distribution.

With regard to the mass-redshift dependence, it can be observed that a qualitative selection function is viewed in the identified clusters. Found objects do not descend at increasing redshift to small masses, which is related to the sensitivity of the SZcat catalogue, but at the same time we find low-mass clusters at small z. In order to recover selection function, it would be necessary to simulate clusters, which is beyond the scope of this paper but will be done in the future.

\section{Conclusions}
\label{sec:conclusions}
Galaxy clusters are the most massive gravitationally bound systems, consisting of dark matter, hot baryonic gas, and stars. They play an important role in observational cosmology and studies of galaxy evolution. We have developed a deep learning model for the segmentation of Sunyaev-Zeldovich (SZ) signals on ACT+Planck intensity maps and constructed a pipeline for the detection of microwave clusters in the ACT footprint. The proposed model allows us to identify previously unknown galaxy clusters, i.e. it is capable of detecting SZ sources below the detection threshold adopted in the published galaxy cluster catalogues (such as ACT DR5 and PSZ2). 

In this paper, we use the derived SZ signal map to considerably improve cluster purity in the extended catalogue of Sunyaev-Zeldovich objects from Planck data (SZcat) in the ACT footprint. From SZcat, we create a new microwave galaxy cluster catalogue (ComPACT), which contains 2,962 SZ-objects with cluster purity conservatively estimated as $\gtrsim74-84$\%. We categorise objects in the catalogue into 3 categories according to their cluster reliability. The ComPACT catalogue includes more than $977$ new clusters (with respect to ACT DR5 and PSZ2 catalogues).

In future work, we plan to identify these objects in Optical and X-ray surveys and estimate their total masses and redshifts.

\section*{Acknowledgements}
We acknowledge the publicly available software packages that were used throughout this work: NumPy \citep{Oliphant:06, Walt:11, Harris:20}, pandas \citep{pandas:2023, mckinney-proc-scipy-2010}, Matplotlib \citep{Hunter:2007}, Astropy \citep{astropy13, astropy_2018, astropy_2022}, pixell\footnote{https://github.com/simonsobs/pixell} and pytorch \citep{Pytorch}. 

We also acknowledge the use of the Legacy Archive for Microwave Background Data Analysis (LAMBDA).

\section*{Data availability}
The ComPACT cluster catalogue is publicly available: \href{https://github.com/astromining/ComPACT}{https://github.com/astromining/ComPACT}
\medskip

\bibliographystyle{mnras}
\bibliography{pcluster} 

\appendix
\thispagestyle{empty}
\section{Description of the ComPACT catalogue}
\begin{adjustbox}{width=\textwidth,totalheight=\textheight,keepaspectratio,rotate=90,caption={Sample of 5 rows from ComPACT}, float=table, label={table:compact} }
    \begin{tabular}{ccccccccc} 
         \hline
         Name & RA & DEC & S & pmax & SZcat & ACT & PSZ2& Priority\\
         \hline\hline
         ComPACT\textunderscore G68.690-52.228&343.02500&-2.40000&27&0.987531&SZ\textunderscore G068.57-52.25&-&-&2\\
         ComPACT\textunderscore G64.587-58.815&346.39167&-8.52500&27&0.840159&SZ\textunderscore G064.59-58.75&-&-&2\\
         ComPACT\textunderscore G7.590-33.933&306.925&-34.79167&59&0.998711&SZ\textunderscore G007.56-33.94&ACT-CL J2027.6-3447&PSZ2 G007.57-33.90&1\\
         ComPACT\textunderscore G32.31337.034&248.78333&15.48333&77&0.999866&SZ\textunderscore G032.32+37.03&ACT-CL J1635.1+1529&-&1\\
         ComPACT\textunderscore G208.402-28.432&75.408333&-8.93333&73&0.996167&SZ\textunderscore G208.45-28.38&-&-&1\\
         \hline
    \end{tabular}
\end{adjustbox}

The ComPACT catalogue contains 2,962 galaxy cluster candidates ($Purity = 74-84$\%). Below we describe columns of the catalogue.
\begin{description}
    \item[\bf Name :] ID of a ComPACT object;
    \item[\bf RA :] Object right ascension (in degrees, J2000 epoch). Centre of object corresponds to pixel with highest SZ signal probability in the source area;
    \item[\bf DEC :] Object declination (in degrees, J2000 epoch);
    \item[\bf S :] Object area (pixels with $p>0.3$) on the SZ signal segmentation map (in pixels);
    \item[\bf pmax :] SZ signal probability (in the object centre);
    \item[\bf SZcat :] SZcat object name;
    \item[\bf ACT :] ACT DR5 cluster name;
    \item[\bf PSZ2 :] PSZ2 cluster name; 
    \item[\bf Priority :] reliability of objects: 
        \begin{itemize}
        \item 1 --- the object has $S > 30$ and $p_{max} = 0.8$, $Purity \sim 0.84 - 0.95$;
        \item 2 --- $S > 25$ and $p_{max} = 0.8$ and $Purity \sim 0.78 - 0.94$;
        \item 3 --- $S > 20$ and $p_{max} = 0.8$ and $Purity \sim 0.74 - 0.92$.
    \end{itemize}  
\end{description}

\section{Description of external cluster catalogues}
\label{sec:appB}
\begin{figure*}
\centering
\includegraphics[width=0.9\textwidth]{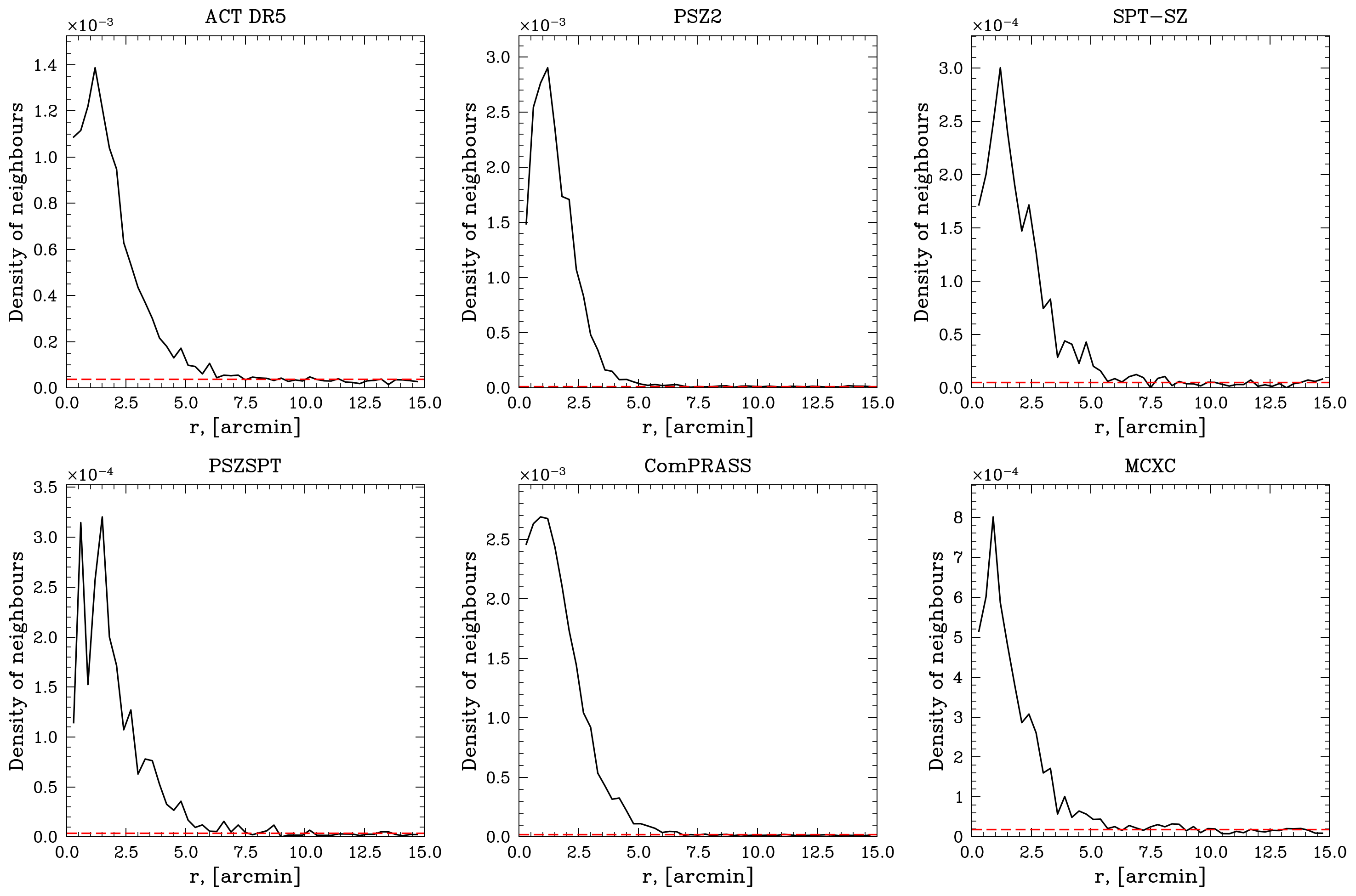}
\caption{Average number density of clusters (from different catalogues) as a function of radius to SZcat sources. The dashed horizontal line shows the mean density.}
\label{fig:r}
\end{figure*}

Below is a description of the catalogues used, which were not described in the main part of the article
\subsection{PSZ2}
The Planck Sunyaev-Zeldovich 2 (PSZ2, \cite{psz2}) cluster catalogue contains clusters with a high level of significance ($S/N > 4.5$). The total number of objects is 1,653. The PSZ2 catalogue includes galaxy clusters with redshifts up to z = 0.8. The multi-frequency matched filter algorithm (MMF1 \cite{Herranz_02:mmf1}, MMF3 \cite{MMF_Melin:06}) and PowellSnakes \cite{PowellSnakes} were used to find SZ objects. In the ACT footprint, there are 782 clusters. Among those, 551 clusters  have also measurements of spectroscopic redshifts. We call this subsample of the Planck catalogue as PSZ2z.  Cluster masses in the PSZ2 catalogue were obtained from the scaling relation between the measured SZ flux, $Y_{500}$, and the total mass $M_{500}$ at a given redshift \citep[for details, see][]{2014A&A...571A..20P}.

\subsection{SPT-SZ}
The South Pole Telescope (SPT) observes the southern sky at 95, 150, and 220 GHz with an angular resolution of 1-1.6 minutes \citep{SPT_Carlstrom_2011}. We use the SPT-SZ cluster catalogue\footnote{\url{https://lambda.gsfc.nasa.gov/product/spt/spt_sz_cluster_catalogs_info.html}} \citep{Bleem_2015, SPT-SZ}, which contains 677 SZ detections, among which 516 objects have been confirmed as clusters using optical and near-infrared observations. 597 SPT-SZ clusters lie in the ACT footprint. Mass estimates for the SPT-SZ clusters come from the weak-lensing calibrated scaling relations \citep{SPT-SZ}.

\subsection{PSZSPT}
The PSZSPT \citep{PSZSPT} catalogue combines data from the Planck Space Telescope and the SPT ground-based telescope. The catalogue was obtained by using multi-frequency matched filtering and contains 419 clusters (with $S/N > 5$), 373 of which lie in the ACT footprint.

\subsection{ComPRASS}
Planck microwave data and RASS (ROSAT All-Sky Survey, \cite{ComPRASS}) X-ray data were combined to create the catalogue. The catalogue contains 2,323 objects (including confirmed clusters and cluster candidates), with 1,110 objects are in the ACT footprint.

\subsection{MCXC}
The MCXC meta catalogue \citep{mcxc} contains 1,868 X-ray clusters detected in ROSAT and EXOSAT data (by various groups) with optical identifications. 867 X-ray clusters are in the ACT footprint.

\section{Dependence between R and S}
\label{sec:appC}

\begin{figure*}
\centering
\includegraphics[width=1\textwidth]{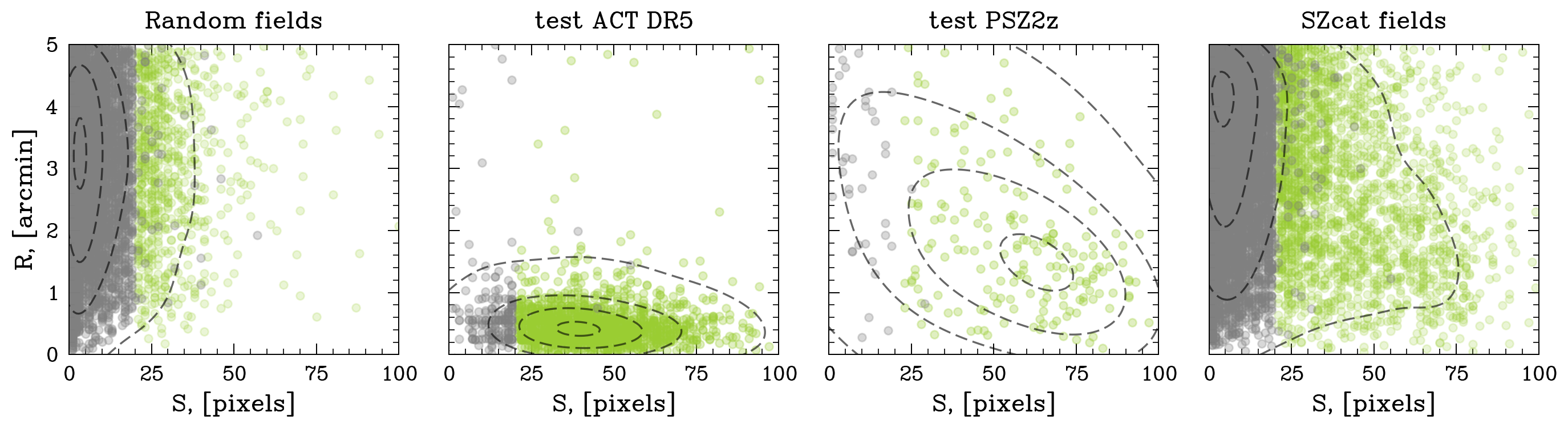}
\caption{Distribution of connected groups of pixels according to their distance $R$ from the input direction and their area $S$. The left panel shows results for random directions that are unlikely to contain galaxy clusters. Density contours are shown as black dashed curves. Two central panels illustrate detected groups of pixels which corresponding to clusters from the ACT DR5 test subsample and spectroscopically confirmed clusters from PSZ2. The right panel shows the distribution of detected groups of pixels for SZcat directions. The latter plot combines characteristics of random fields and PSZ2 clusters since SZcat contains a quite number of false detections. In green, we show groups that satisfy our criteria for clusters: $S>20$ and $p_{max}>0.8$.}  
\label{fig:quan}
\end{figure*}

Let us now illustrate how a distribution of detected sources looks like on the $R$ -- $S$ plane for random directions\footnote{but within the ACT field} on the sky (see the left panel of Fig.~\ref{fig:quan}). Random fields are unlikely to contain galaxy clusters. 

We see that non-clusters tend to have low areas and are offset from the input direction. In central panels Fig.~\ref{fig:quan} we show distributions of clusters detected by our model along the input directions from the ACT DR5 test subsample (see Section~\ref{sec:dataset}) and from PSZ2z (the PSZ2 subsample of clusters with spectroscopic redshifts). We see that real clusters tend to have large areas. For ACT DR5 sources, $R$ is low ($\sim$ 1 arcmin for majority of clusters) since ACT has a good spatial resolution. For PSZ2z, clusters are more widely distributed in the $R$--$S$ plane because of poorer Planck resolution (compared to ACT). Right panel of Fig.~\ref{fig:quan} present distributions of nearest connected groups of pixels detected by our model in the 5 arcmin window along the SZcat directions. We do not use R as a threshold because it varies from catalogue to catalogue FWHM.

\section{Catalogue metrics}
\label{app:metrics}

The key characteristics of a cluster catalogue are purity and completeness metrics. 

\begin{enumerate}
    \item Completeness, which we define as 
        \begin{equation}
            C=\cfrac{N_{cl,thresh}}{N_{cl}} ~,
        \label{eq:completness}
        \end{equation}
        where $N_{cl,thresh}$ is the total number of galaxy clusters with chosen threshold on $p_{max}$ and $S$, $N_{cl}$ is the total number of clusters in the considered catalogue;
    \item We define the purity of the ComPACT cluster catalogue as: 
        \begin{equation}
            Purity=1 - \cfrac{N_{F,ComPACT}}{N_{ComPACT}} ~,
        \label{eq:purity0}
        \end{equation}
    where $N_{F,ComPACT}$ and $N_{ComPACT}$ are the number of false detections and the total number of objects in the ComPACT catalogue, respectively. The total number of galaxy clusters in the ComPACT catalogue ($N_{cl,ComPACT}=N_{ComPACT}-N_{F,ComPACT}$) can be divided into 3 parts:
    \begin{itemize}
        \item Planck clusters ($N_{cl,Planck}$). SZ objects seen in Planck data, which correspond to excess of objects in ComPACT catalogue with respect to random fields directions:
        \begin{equation}
            N_{cl,Planck} = N_{ComPACT} - \hat{N}_{field} ~,
        \end{equation}
        where $\hat{N}_{field}$ --- the number of detections ($p_{max}>0.8$, $S>20$) in the random fields directions ($N_{field}=N_{SZcat}$, see \S\ref{sec:dataset}). One can obtain the following estimate of ComPACT purity with respect to Planck clusters:
        \begin{equation}
            Purity_{cl,Planck} = 1 - \cfrac{\hat{N}_{field}}{N_{ComPACT}} ~.
        \label{eq:purity_planck}
        \end{equation}
    \item ACT clusters ($N_{cl,ACT}$). SZ objects, which are seen in ACT data only. The lower bound estimate of this number one can obtain as follows. According to \citet{SZcat}, only 21.7\% SZcat objects are associated with eROSITa clusters (1-year all-sky survey data) on the Eastern galactic hemisphere. Thus for non-Planck clusters we have:
    \begin{equation}
        Purity_{ACT} \approx \cfrac{0.217\cdot{}N_{SZcat}-N_{PSZ2}}{N_{SZcat}-N_{PSZ2}}  ,
    \end{equation}
    where $N_{PSZ2}$ means number of objects in the ComPACT catalogue associated with PSZ2. ComPACT purity:
    \begin{equation}
        Purity_{min} \approx \cfrac{N_{cl, Planck} + Purity_{ACT}\hat{N}_{field}}{N_{ComPACT}} ~.
    \label{eq:purity_min}
    \end{equation}

    \item The number of false detections can be estimated as 
        \begin{equation}
            N_{F}\lesssim{}(1 - P)\cdot{}\hat{N}_{field} ~,
        \label{eq:N_F}
        \end{equation}
        where $P=0.8$ is our threshold in the maximum probability ($p_{max}$). The idea behind formula (\ref{eq:N_F}) is simple. The subsample of detections in random fields with $p_{max}>0.8$ and $S>20$ contains more than $ P\times 100\% = 80\%$ of galaxy clusters. Thus, the number of false detections in this subsample can be estimated as  $<(1 - P)\cdot{}\hat{N}_{field}$.
        Based on the expressions (\ref{eq:purity0}) and (\ref{eq:N_F}) one can obtain:
        \begin{equation}
            Purity_{high} = 1 - (1 - P)\cdot\bigg(\cfrac{\hat{N}_{field}}{N_{ComPACT}}\bigg)  ~.
        \label{eq:purity_high}
        \end{equation}
    \end{itemize}

\end{enumerate}

\bsp	
\label{lastpage}
\end{document}